\documentclass[12pt]{article}
\usepackage{amsmath}
\usepackage{amssymb,scalefnt}
\usepackage{amsfonts}
\usepackage{latexsym}
\usepackage{color}
\usepackage{dsfont}
\usepackage{graphicx}
\usepackage{epstopdf}
\usepackage{pdfpages}

\catcode `\@=11 \@addtoreset{equation}{section}

\catcode `\@=12
\newcommand{\be}{\begin{equation}}
\newcommand{\en}{\end{equation}}
\newcommand{\bea}{\begin{eqnarray}}
\newcommand{\ena}{\end{eqnarray}}
\newcommand{\beano}{\begin{eqnarray*}}
\newcommand{\enano}{\end{eqnarray*}}
\newcommand{\bee}{\begin{enumerate}}
\newcommand{\ene}{\end{enumerate}}

\newcommand{\R}{\mathbb{R}}
\newcommand{\mc}{\mathcal}

\newcommand{\Pc}{{\cal P}}
\newcommand{\Rc}{{\cal R}}
\newcommand{\Sc}{{\cal S}}

\newcommand{\F}{{\cal F}}

\newcommand{\1}{1 \!\! 1}

\newcommand{\Hil}{\mc H}

\catcode `\@=11 \@addtoreset{equation}{section}
\catcode `\@=12
\begin{document}

\begin{center}
{\Large \textbf{An improved model of
alliances between political parties}}\vspace{1.5cm}%
\\[0pt]

{\large F. Bagarello}\\[0pt]
Dipartimento di Energia, Ingegneria dell'Informazione e Modelli Matematici,\\%
[0pt]
Scuola Politecnica Ingegneria, Universit\`a di Palermo,\\[0pt]
I-90128 Palermo, Italy\\[0pt]
and I.N.F.N., Sezione di Torino\\[0pt]
e-mail: fabio.bagarello@unipa.it\\[0pt]
home page: www.unipa.it/fabio.bagarello

\end{center}

\vspace*{.5cm}

\begin{abstract}
\noindent We consider an  operatorial model of alliances between three political parties which interact with their electors, with the undecided voters, and with the electors of the other parties. This extends what was done in a previous paper, where this last type of interactions was not considered. Of course, taking them into account makes the system closer to real life. To produce an exactly solvable model, we restrict here to quadratic Hamiltonians, so that the equations of motion turn out to be linear. The dynamics of the so-called {\em decision functions} are deduced, and some explicit situations are considered in details.

\end{abstract}

\thispagestyle{empty}


\vspace{2cm}


\vfill


\newpage

\section{Introduction}

In recent years, several models of {\em decision making processes} have been introduced and analyzed in the mathematical and physical literature, by many authors and in different contexts. In particular, quantum approaches to decision making have been proposed in \cite{qdm1}-\cite{qdm5}. However, very few attempts exist connecting decision making and politics, and, in particular,  trying to model alliances in politics. Some mathematical approaches to politics are discussed in \cite{pol1}-\cite{pol5} and in references therein, but none of them deals with this particular problem.

In a series of recent papers, \cite{all1}-\cite{all3}, a model of interaction between political
parties has been proposed and analyzed in several versions. Each model describes a decision making
procedure, deducing the time evolution of three so-called \emph{decision
functions} (DFs), one for each part{y} considered in our system.
These functions describe the interest of each party to form or not an
alliance with some other party. In particular, in \cite{all1,all2} these decisions are driven by the
interaction of each party with the other parties, with their own electors,
and with a set of undecided voters (i.e. people who have not yet decided  for which party to
vote (if at all they decide to vote!)). The approach
adopted in \cite{all1} uses an operatorial framework {(see also} \cite%
{bagbook}), in which the DFs are suitable mean values of certain number
operators associated to the parties. The dynamics {is} driven by a
suitable Hamiltonian which implements the various interactions between the
different actors of the system. In \cite{all3} no set of electors is considered at all, and we rather focus on the interactions between the three parties, interactions which are taken to be of different forms, all giving rise to exact solutions.

Our three papers cited above differ in several aspects: \cite{all1} and \cite{all2} share the same general features, with parties and electors giving rise to a sort of {\em open system}. In \cite{all1} the equations of motion are linear and oversimplified, while in \cite{all2} we consider nonlinear interactions which, however, are so complicated that  no exact analytical (or even numerical) solution is possible: only approximated solutions can be deduced. In \cite{all3}, on the other hand, paying the price to neglect the interactions between the parties and the electors, we are able to find, easily enough, the exact expressions of the DFs for the three parties, even in presence of nonlinearities: now we have a simpler {\em closed system}, having a finite number of degrees of freedom. Not surprisingly, none of the models proposed so far is {\em complete}. In fact, completeness of the model is deeply related with the difficulty of deducing its (dynamical) solution. Still, there is something we can do to extend the original model in \cite{all1} to make it more realistic, while keeping the analytical difficulties at a reasonable level. The extension we will discuss in this paper goes like this: while, as already mentioned, in \cite{all1} and \cite{all2}  the three parties $\Pc_1$, $\Pc_2$ and $\Pc_3$ can only interact with their respective electors $\Rc_1$, $\Rc_2$ and $\Rc_3$, and with the undecided voters ($\Rc_{und}$), here we  consider the possibility that, for instance, $\Pc_1$  interacts also with $\Rc_2$ and with $\Rc_3$, and so on.

The kind of interaction that we will adopt here is quadratic, and this gives rise to exactly solvable equations of motion, as we will see. The paper is organized as follows: in the next section we will briefly recall the model in \cite{all1}. This is useful to introduce the notation and clarify the approach. Section \ref{sectIII} contains our extension, the differential equations which are derived by the new Hamiltonian and the solution of these equations. Then, in Section \ref{sectIV}, we consider briefly a particular situation, obtained by fixing some specific values of the parameters of the Hamiltonian and certain initial conditions, and we  deduce the time evolution of the DFs for the three parties. Section \ref{sectV} contains our conclusions.

\section{The original model}

In our model we have three parties, $\Pc_1$, $\Pc_2$ and $\Pc_3$, which, together, form the system $\Sc_\Pc$. Each party has to make a choice, and it can only choose one or zero, corresponding respectively to {\em form a coalition} with some other party or not.  Hence we have eight different possibilities, which we associate to eight different and mutually orthogonal vectors in an eight-dimensional Hilbert space $\Hil_\Pc$. These vectors are called $\varphi_{i,k,l}$, with $i,k,l=0,1$. The three subscripts refers to whether or not the three parties of the models wants to form a coalition at time $t=0$. Hence, for example, the vector $\varphi_{0,0,0}$, describes the fact that, at $t=0$, no party wants to ally with the other parties. Of course, this attitude can change during the time evolution, and deducing these changes is, in fact, what is interesting for us.   The set $\F_\varphi=\{\varphi_{i,k,l},\,i,k,l=0,1\}$ is an orthonormal basis for $\Hil_\Pc$.

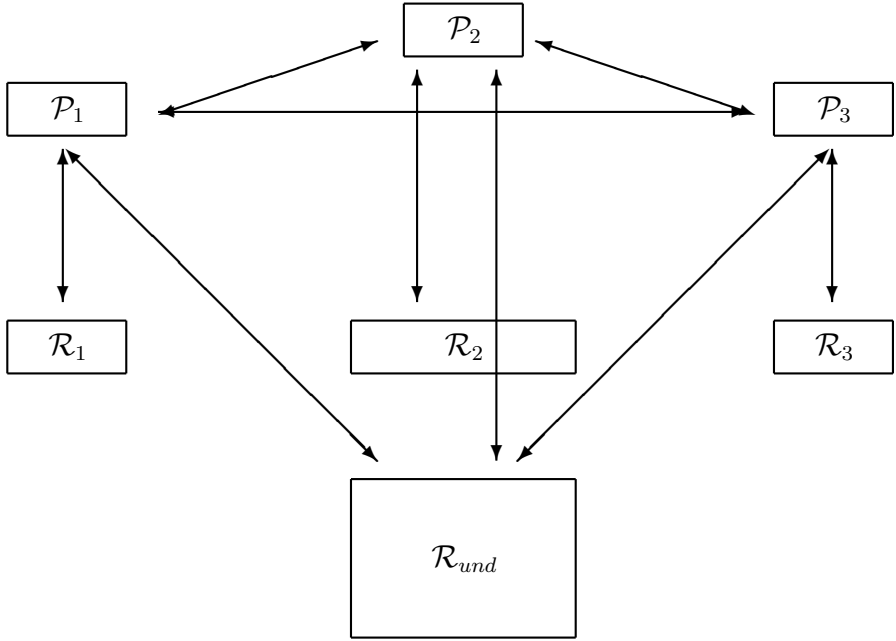
\begin{figure}
\begin{center}
\begin{picture}(450,90)

\put(160,65){\thicklines\line(1,0){45}}
\put(160,85){\thicklines\line(1,0){45}}
\put(160,65){\thicklines\line(0,1){20}}
\put(205,65){\thicklines\line(0,1){20}}
\put(183,75){\makebox(0,0){$\Pc_2$}}

\put(300,35){\thicklines\line(1,0){45}}
\put(300,55){\thicklines\line(1,0){45}}
\put(300,35){\thicklines\line(0,1){20}}
\put(345,35){\thicklines\line(0,1){20}}
\put(323,45){\makebox(0,0){$\Pc_3$}}

\put(10,35){\thicklines\line(1,0){45}}
\put(10,55){\thicklines\line(1,0){45}}
\put(10,35){\thicklines\line(0,1){20}}
\put(55,35){\thicklines\line(0,1){20}}
\put(33,45){\makebox(0,0){$\Pc_1$}}

\put(10,-55){\thicklines\line(1,0){45}}
\put(10,-35){\thicklines\line(1,0){45}}
\put(10,-55){\thicklines\line(0,1){20}}
\put(55,-55){\thicklines\line(0,1){20}}
\put(33,-45){\makebox(0,0){$\Rc_1$}}

\put(140,-55){\thicklines\line(1,0){85}}
\put(140,-35){\thicklines\line(1,0){85}}
\put(140,-55){\thicklines\line(0,1){20}}
\put(225,-55){\thicklines\line(0,1){20}}
\put(183,-45){\makebox(0,0){$\Rc_2$}}

\put(300,-55){\thicklines\line(1,0){45}}
\put(300,-35){\thicklines\line(1,0){45}}
\put(300,-55){\thicklines\line(0,1){20}}
\put(345,-55){\thicklines\line(0,1){20}}
\put(323,-45){\makebox(0,0){$\Rc_3$}}

\put(140,-155){\thicklines\line(1,0){85}}
\put(140,-95){\thicklines\line(1,0){85}}
\put(140,-155){\thicklines\line(0,1){60}}
\put(225,-155){\thicklines\line(0,1){60}}
\put(183,-125){\makebox(0,0){$\Rc_{und}$}}

\put(70,44){\thicklines\vector(1,0){220}}
\put(70,44){\thicklines\vector(-1,0){3}}
\put(70,44){\thicklines\vector(3,1){80}}
\put(70,44){\thicklines\vector(-3,-1){3}}
\put(290,44){\thicklines\vector(-3,1){80}}
\put(290,44){\thicklines\vector(3,-1){3}}

\put(31,27){\thicklines\vector(0,-1){55}}
\put(31,27){\thicklines\vector(0,1){3}}
\put(322,27){\thicklines\vector(0,-1){55}}
\put(322,27){\thicklines\vector(0,1){3}}
\put(165,57){\thicklines\vector(0,-1){85}}
\put(165,57){\thicklines\vector(0,1){3}}

\put(35,27){\thicklines\vector(1,-1){115}}
\put(35,27){\thicklines\vector(-1,1){3}}
\put(318,27){\thicklines\vector(-1,-1){115}}
\put(318,27){\thicklines\vector(1,1){3}}
\put(195,57){\thicklines\vector(0,-1){145}}
\put(195,57){\thicklines\vector(0,1){3}}


\end{picture}
 \end{center}
\vspace*{5.3cm}
\caption{\label{figscheme} The original system and its multi-component reservoir.}
\end{figure}

As we have shown in \cite{all1}, and later on in \cite{all2,all3}, it is convenient to construct the vectors $\varphi_{i,k,l}$ in a very special way, starting with the vacuum of three fermionic operators, $p_1$, $p_2$ and $p_3$, i.e. three operators which, together with their adjoint, satisfy the canonical anticommutation relation (CAR) $\{p_k,p_l^\dagger\}=\delta_{k,l}$ and $\{p_k,p_l\}=0$. Then,  $\varphi_{0,0,0}$ is a vector satisfying $p_j\varphi_{0,0,0}=0$, $j=1,2,3$, and the other vectors $\varphi_{i,k,l}$  can be constructed
 out of $\varphi_{0,0,0}$ as follows:
$$
\varphi_{1,0,0}=p_1^\dagger\varphi_{0,0,0}, \quad \varphi_{0,1,0}=p_2^\dagger\varphi_{0,0,0}, \quad \varphi_{1,1,0}=p_1^\dagger\,p_2^\dagger\varphi_{0,0,0},\quad \varphi_{1,1,1}=p_1^\dagger\,p_2^\dagger\,p_3^\dagger\varphi_{0,0,0},
$$
and so on. Let now $\hat P_j=p_j^\dagger p_j$ be the so-called {\em number operator} of the $j$-th party, which is constructed using $p_j$ and its adjoint, $p_j^\dagger$. Since $\hat P_j\varphi_{n_1,n_2,n_3}=n_j\varphi_{n_1,n_2,n_3}$, for $j=1,2,3$, the eigenvalues of these operators, zero and one, correspond to the only possible choices of the three parties at $t=0$. This is, in fact, the main reason why we have used here the fermionic operators $p_j$: they automatically produce only these eigenvalues. We have seen in \cite{all1}-\cite{all3} how to give a dynamics to the number operator $\hat P_j$, following the general scheme described in \cite{bagbook} and adopted in very different contexts. In this way, we can follow how the parties modify their attitude with respect to time, regarding alliances. This is achieved by fixing, first of all, a suitable Hamiltonian, which describes the interactions indicated by the arrows in Figure \ref{figscheme}. Here $\Rc_j$ represents the set of the supporters of $\Pc_j$, while $\Rc_{und}$ is the set of all the undecided electors. This figure also shows that, for instance, $\Pc_1$ can interact with $\Pc_2$ and $\Pc_3$, and with $\Rc_1$ and $\Rc_{und}$, but not with $\Rc_2$ or with $\Rc_3$. This is exactly the limitation that, in the next section, we will remove.

 \vspace*{1cm}

The Hamiltonian which describes  the scheme in Figure \ref{figscheme},  written in terms of fermionic operators, is the following, \cite{all1}:
\be
\left\{
\begin{array}{ll}
H=H_{0}+H_{PBs}+H_{PB}+H_{int}, &  \\
H_{0}=\sum_{j=1}^{3}\omega _{j}p_j^\dagger p_j+\sum_{j=1}^{3}\int_{\Bbb R}\Omega_j(k)B_j^\dagger(k)B_j(k)\,dk+\int_{\Bbb R}\Omega(k)B^\dagger(k)B(k)\,dk,   \\
H_{PBs}=\sum_{j=1}^{3}\lambda_j\int_{\Bbb R}\left(p_j B_j^\dagger(k)+B_j(k)p_j^\dagger\right)\,dk,\\
H_{PB}=\sum_{j=1}^{3}\tilde\lambda_j\int_{\Bbb R}\left(p_j B^\dagger(k)+B(k)p_j^\dagger\right)\,dk,\\
H_{int}=\mu_{12}^{ex}\left(p_1^\dagger p_2+p_2^\dagger p_1\right)+\mu_{12}^{coop}\left(p_1^\dagger p_2^\dagger+p_2 p_1\right)+\mu_{13}^{ex}\left(p_1^\dagger p_3+p_3^\dagger p_1\right)+   \\
\qquad +\mu_{13}^{coop}\left(p_1^\dagger p_3^\dagger+p_3 p_1\right)+\mu_{23}^{ex}\left(p_2^\dagger p_3+p_3^\dagger p_2\right)+\mu_{23}^{coop}\left(p_2^\dagger p_3^\dagger+p_3 p_2\right).
\end{array}%
\right.
\label{22}\en
Here $\omega _{j}$, $\lambda_j$, $\tilde\lambda_j$, $\mu_{ij}^{ex}$ and $\mu_{ij}^{coop}$ are real quantities, while $\Omega_j(k)$ and $\Omega(k)$ are real-valued functions, whose meaning is explained in \cite{all1}, together with the meaning of each term of $H$.  Here we just want to stress  that the three parties are considered as a part of a larger system: in order to take their decisions, they need first to interact with the electors and among themselves, since it is exactly this interaction which motivates their final decisions. Hence, $\Sc_\Pc$ must be {\em open}, i.e. there must be some environment, $\Rc$ (the full set of electors),  interacting with $\Pc_1$, $\Pc_2$ and $\Pc_3$, which produces some  feedback used by $\Pc_j$ to decide what to do. Moreover, the environment, when compared with $\Sc_\Pc$, is expected to be very large, since the sets of the electors for $\Pc_1$, $\Pc_2$ and $\Pc_3$ are supposed to be sufficiently large. This is the reason why (infinitely many) operators $B_i(k)$ and $B_i^\dagger(k)$, $k\in\Bbb R$, appear in $H$.

 The following CAR's for the operators of the reservoir are assumed:
\be
\{B_i(k),B_l^\dagger(q)\}=\delta_{i,l}\delta(k-q)\,\1,\qquad \{B_i(k),B_l(k)\}=0,
\label{23}
\en
as well as
\be
\{B(k),B^\dagger(q)\}=\delta(k-q)\,\1,\quad \{B(k),B(k)\}=0,
\label{23b}
\en
for all $i,l=1,2,3$, $k,\, q\in{\Bbb R}$. Moreover each $p_j^\sharp$ anti-commutes with each $B_l^\sharp(k)$ and with $B^\sharp(k)$: $\{p_j^\sharp, B_l^\sharp(k)\}=\{p_j^\sharp, B^\sharp(k)\}=0$ for all $j$, $l$ and for all $k$, and we further assume that $\{B^\sharp(q), B_l^\sharp(k)\}=0$, for all $k,q\in\Bbb R$. Here $X^\sharp$ stands for $X$ or $X^\dagger$. Assuming these CAR's is natural, since they reflect the analogous choice adopted for the operators of the three parties.

Once $H$ is given, we have to
compute the time evolution of the number operators in the Heisenberg scheme as $\hat P_j(t):=e^{iHt}\hat P_j e^{-iHt}$, and then their mean values on some suitable state describing the full system (parties and electors) at $t=0$. This is what in \cite{all1}-\cite{all3} has been called {\em decision function}, see formula (\ref{add1}) below.

\vspace{1mm}

We can now go back to the analysis of the dynamics of the system. The Heisenberg equations of motion $\dot X(t)=i[H,X(t)]$, \cite{bagbook}, can be deduced by using the CAR's (\ref{23}) and (\ref{23b}) above. In \cite{all1} we have deduced the following set of equations, using the operator $H$ in (\ref{22}):

\be
\left\{
\begin{array}{ll}
\dot p_1(t)=-i\omega_1 p_1(t)+i\lambda_1\int_{\Bbb R}B_1(q,t)\,dq+i\tilde\lambda_1\int_{\Bbb R}B(q,t)\,dq-i\mu_{12}^{ex}p_2(t)-i\mu_{12}^{coop}p_2^\dagger(t)+\\
\qquad -i\mu_{13}^{ex}p_3(t)-i\mu_{13}^{coop}p_3^\dagger(t),   \\
\vspace{1mm}
\dot p_2(t)=-i\omega_2 p_2(t)+i\lambda_2\int_{\Bbb R}B_2(q,t)\,dq+i\tilde\lambda_2\int_{\Bbb R}B(q,t)\,dq-i\mu_{12}^{ex}p_1(t)+i\mu_{12}^{coop}p_1^\dagger(t)+\\
\qquad -i\mu_{23}^{ex}p_3(t)-i\mu_{23}^{coop}p_3^\dagger(t),   \\
\vspace{1mm}
\dot p_3(t)=-i\omega_3 p_3(t)+i\lambda_3\int_{\Bbb R}B_3(q,t)\,dq+i\tilde\lambda_3\int_{\Bbb R}B(q,t)\,dq-i\mu_{13}^{ex}p_1(t)+i\mu_{13}^{coop}p_1^\dagger(t)+\\
\qquad -i\mu_{23}^{ex}p_2(t)+i\mu_{23}^{coop}p_2^\dagger(t),   \\
\vspace{1mm}
\dot B_j(q,t)=-i\Omega_j(q) B_j(q,t)+i\lambda_j p_j(t),\qquad j=1,2,3,   \\
\vspace{1mm}
\dot B(q,t)=-i\Omega(q) B(q,t)+i\sum_{j=1}^3\tilde\lambda_j p_j(t).   \label{26}
\end{array}%
\right.
\en
These equations are solved and $p_1(t)$, $p_2(t)$ and $p_3(t)$ are deduced. In this way  the number operators $\hat P_j(t)=p_j^\dagger(t)p_j(t)$, $j=1,2,3$, are found, and the DFs are obtained as  \be P_j(t):=\left<\hat P_j(t)\right>=\left<p_j^\dagger(t)p_j(t)\right>,\label{add1}\en $j=1,2,3$. Here $\left<.\right>$ is a state over the full system. These states, \cite{bagbook}, are taken to be suitable tensor products of vector states on $\Sc_\Pc$ and states on the reservoir which obey some standard rules, see below. More in details,  for each operator of the form $X_{\Sc}\otimes Y_{\Rc}$, $X_{\Sc}$ being an operator of $\Sc_\Pc$ and $Y_{\Rc}$ an operator of
the reservoir, we put
\be
\left\langle X_{\Sc}\otimes Y_{\Rc}\right\rangle :=\left\langle \varphi_{n_1,n_2,n_3},X_{\Sc}\varphi_{n_1,n_2,n_3}\right\rangle \,\omega
_{\Rc}(Y_{\Rc}).
\label{add2}\en
Here $\varphi_{n_1,n_2,n_3}$ is, as already stated, one of the vectors introduced at the beginning of this section, and each $n_j$ represents, as discussed before, the tendency of $\Pc_j$ to form or not some coalition at $t=0$. Moreover, $\omega _{\Rc}(.)$ is a state on $\R$ satisfying
the following standard properties, \cite{bagbook}:
\be
\omega _{\Rc}(1\!\!1_{\Rc})=1,\quad \omega _{\Rc}(B_{j}(k))=\omega
_{\Rc}(B_{j}^{\dagger }(k))=0,\quad \omega _{\Rc}(B_{j}^{\dagger
}(k)B_{l}(q))=N_{j}(k)\,\delta _{j,l}\delta (k-q),
\label{211}\en
as well as
\be
\omega _{\Rc}(B(k))=\omega
_{\Rc}(B^{\dagger }(k))=0,\quad \omega _{\Rc}(B^{\dagger
}(k)B(q))=N(k)\,\delta (k-q),
\label{211bis}\en
for some suitable functions $N_{j}(k)$, $N(k)$ which we take here to be constant in $k$: $N_{j}(k)=N_j$ and $N(k)=N$.  Also, we assume that $\omega
_{\Rc}(B_{j}(k)B_{l}(q))=\omega
_{\Rc}(B(k)B(q))=0$, for all $j, l=1,2,3$, and for all $k, q\in\Bbb R$. In our framework, the state in (\ref{add2}) describes the fact that, at $t=0$,  $\Pc_j$'s decision (concerning alliances) is $n_j$ ($P_j(0)=n_j$), while the overall feeling of the voters $\Rc_j$ is $N_j$, and that of the undecided ones is $N$. Of course, these might appear as oversimplifying assumptions, and in fact they are. However, they still produce, in many concrete applications, a rather interesting dynamics for the model.

We refer to \cite{all1} for several explicit situations described by (\ref{26}). Here we are more interested in considering what happens when the original scheme in Figure \ref{figscheme} is extended as discussed in the next section.

\section{The extended model}\label{sectIII}

As already stated, we are more interested in considering possible interactions not only between $\Pc_j$ and $\Rc_j$, but also between $\Pc_j$ and $\Rc_k$, with $j\neq k$. In other words, we want to see what happens if the various electors can talk with each party, and how their DFs are modified because of this possibility.

\begin{figure}
\begin{center}
\begin{picture}(450,90)

\put(160,65){\thicklines\line(1,0){45}}
\put(160,85){\thicklines\line(1,0){45}}
\put(160,65){\thicklines\line(0,1){20}}
\put(205,65){\thicklines\line(0,1){20}}
\put(183,75){\makebox(0,0){$\Pc_2$}}

\put(300,35){\thicklines\line(1,0){45}}
\put(300,55){\thicklines\line(1,0){45}}
\put(300,35){\thicklines\line(0,1){20}}
\put(345,35){\thicklines\line(0,1){20}}
\put(323,45){\makebox(0,0){$\Pc_3$}}

\put(10,35){\thicklines\line(1,0){45}}
\put(10,55){\thicklines\line(1,0){45}}
\put(10,35){\thicklines\line(0,1){20}}
\put(55,35){\thicklines\line(0,1){20}}
\put(33,45){\makebox(0,0){$\Pc_1$}}

\put(10,-55){\thicklines\line(1,0){45}}
\put(10,-35){\thicklines\line(1,0){45}}
\put(10,-55){\thicklines\line(0,1){20}}
\put(55,-55){\thicklines\line(0,1){20}}
\put(33,-45){\makebox(0,0){$\Rc_1$}}

\put(140,-55){\thicklines\line(1,0){85}}
\put(140,-35){\thicklines\line(1,0){85}}
\put(140,-55){\thicklines\line(0,1){20}}
\put(225,-55){\thicklines\line(0,1){20}}
\put(183,-45){\makebox(0,0){$\Rc_2$}}

\put(300,-55){\thicklines\line(1,0){45}}
\put(300,-35){\thicklines\line(1,0){45}}
\put(300,-55){\thicklines\line(0,1){20}}
\put(345,-55){\thicklines\line(0,1){20}}
\put(323,-45){\makebox(0,0){$\Rc_3$}}

\put(140,-155){\thicklines\line(1,0){85}}
\put(140,-95){\thicklines\line(1,0){85}}
\put(140,-155){\thicklines\line(0,1){60}}
\put(225,-155){\thicklines\line(0,1){60}}
\put(183,-125){\makebox(0,0){$\Rc_{und}$}}

\put(70,44){\thicklines\vector(1,0){220}}
\put(70,44){\thicklines\vector(-1,0){3}}
\put(70,44){\thicklines\vector(3,1){80}}
\put(70,44){\thicklines\vector(-3,-1){3}}
\put(290,44){\thicklines\vector(-3,1){80}}
\put(290,44){\thicklines\vector(3,-1){3}}

\put(31,27){\thicklines\vector(0,-1){55}}
\put(31,27){\thicklines\vector(0,1){3}}
\put(322,27){\thicklines\vector(0,-1){55}}
\put(322,27){\thicklines\vector(0,1){3}}
\put(161,57){\thicklines\vector(0,-1){85}}
\put(161,57){\thicklines\vector(0,1){3}}

\put(35,27){\thicklines\vector(1,-1){115}}
\put(35,27){\thicklines\vector(-1,1){3}}
\put(318,27){\thicklines\vector(-1,-1){115}}
\put(318,27){\thicklines\vector(1,1){3}}
\put(195,57){\thicklines\vector(0,-1){145}}
\put(195,57){\thicklines\vector(0,1){3}}

\put(35,27){\thicklines\vector(2,-1){105}}
\put(35,27){\thicklines\vector(4,-1){255}}
\put(318,27){\thicklines\vector(-2,-1){105}}
\put(318,27){\thicklines\vector(-4,-1){255}}
\put(195,57){\thicklines\vector(1,-1){95}}
\put(161,58){\thicklines\vector(-1,-1){95}}


\end{picture}
 \end{center}
\vspace*{5.3cm}
\caption{\label{figscheme2} The enriched system and its multi-component reservoir.}
\end{figure}
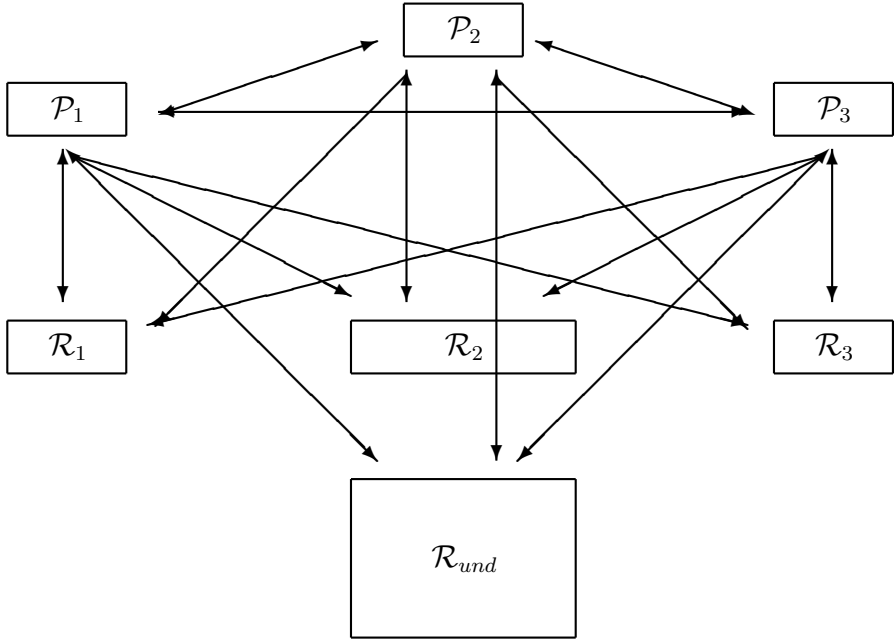

In the case we are interested here Figure \ref{figscheme} must be {\em enriched with more arrows}. In fact, these new arrows are just the new allowed interactions. In particular, the scheme of the system we will describe in this section is given in Figure \ref{figscheme2}. The differences with respect to Figure 1 are clearly given by the arrows connecting $\Pc_j$ with $\Rc_k$, $j\neq k$.

 \vspace*{1cm}

The new Hamiltonian $h$ describing this enriched system can be written as

\be
\left\{
\begin{array}{ll}
&h=H+H_{mix},   \\
&H_{mix}=H_{mix}^p+H_{mix}^{ap},   \\
&H_{mix}^p=\sum_{n\neq l,1}^3\nu_{nl}^p\int_{\Bbb R}dk\left(p_nB_l^\dagger(k)+B_l(k)p_n^\dagger\right),\\
&H_{mix}^{ap}=\sum_{n\neq l,1}^3\nu_{nl}^{ap}\int_{\Bbb R}dk\left(p_n^\dagger B_l^\dagger(k)+B_l(k)p_n\right),\\
\end{array}%
\right.
\label{31}\en
where $H$ is the operator introduced in (\ref{22}), i.e. the original one considered in \cite{all1}. The contribution $H_{mix}$ is, within the scheme adopted here, the operator which describes the new arrows added in Figure \ref{figscheme2}\footnote{In \cite{all2} an extra term has also been added to $H$, to get the full Hamiltonian of the system. However, the nature of the term added in \cite{all2} is completely different from that of $H_{mix}$, since this still produces an analytically solvable set of differential equations.}. In particular, a contribution like $p_nB_l^\dagger(k)$ means that if the party $\Pc_n$ is not willing to form any alliance, then the electors of $\Pc_l$, $\Rc_l$, suggest $\Pc_l$ not to form any alliance either; in fact, their {\em global reaction against alliances} increase, \cite{all1}. Then $\Pc_n$ and $\Pc_l$ tend to have a similar behavior, and this explain the suffix $p$ in  $H_{mix}^p$: $p$ stands for {\em parallel}. In a similar way the term $p_n^\dagger B_l^\dagger(k)$ describes the fact that if the party $\Pc_n$ is now willing to form some alliance, still  $\Rc_l$ suggests  $\Pc_l$ not to form any alliance: $\Pc_n$ and $\Pc_l$ tend to have opposite behaviors, and we use now the suffix $ap$ in  $H_{mix}^{ap}$, which stands for {\em anti-parallel}.

The equations in (\ref{26}) must now be replaced by the system:

\be
\left\{
\begin{array}{ll}
\dot p_1(t)=-i\omega_1 p_1(t)+i\lambda_1\int_{\Bbb R}B_1(q,t)\,dq+i\tilde\lambda_1\int_{\Bbb R}B(q,t)\,dq-i\mu_{12}^{ex}p_2(t)-i\mu_{12}^{coop}p_2^\dagger(t)+\\
\qquad -i\mu_{13}^{ex}p_3(t)-i\mu_{13}^{coop}p_3^\dagger(t)+M_1(t),   \\
\vspace{1mm}
\dot p_2(t)=-i\omega_2 p_2(t)+i\lambda_2\int_{\Bbb R}B_2(q,t)\,dq+i\tilde\lambda_2\int_{\Bbb R}B(q,t)\,dq-i\mu_{12}^{ex}p_1(t)+i\mu_{12}^{coop}p_1^\dagger(t)+\\
\qquad -i\mu_{23}^{ex}p_3(t)-i\mu_{23}^{coop}p_3^\dagger(t)+M_2(t),   \\
\vspace{1mm}
\dot p_3(t)=-i\omega_3 p_3(t)+i\lambda_3\int_{\Bbb R}B_3(q,t)\,dq+i\tilde\lambda_3\int_{\Bbb R}B(q,t)\,dq-i\mu_{13}^{ex}p_1(t)+i\mu_{13}^{coop}p_1^\dagger(t)+\\
\qquad -i\mu_{23}^{ex}p_2(t)+i\mu_{23}^{coop}p_2^\dagger(t)+M_3(t),   \\
\vspace{1mm}
\dot B_j(q,t)=-i\Omega_j(q) B_j(q,t)+i\lambda_j p_j(t)+i\,R_j(t),\qquad\qquad j=1,2,3,   \\
\vspace{1mm}
\dot B(q,t)=-i\Omega(q) B(q,t)+i\sum_{j=1}^3\tilde\lambda_j p_j(t).   \label{32}
\end{array}%
\right.
\en

where we have introduced the following quantities\footnote{Notice that, to simplify the notation, in $M_j$ and $R_j$ we are making explicit only the dependence on $t$.}:

\be
\left\{
\begin{array}{ll}
M_1(t)=i{\int}_{\Bbb R}\left(\nu_{12}^pB_2(q,t)+\nu_{13}^pB_3(q,t)-\nu_{12}^{ap}B_2^\dagger(q,t)-
\nu_{13}^{ap}B_3^\dagger(q,t)\right)dq\\
\vspace{1mm}
M_2(t)=i{\int}_{\Bbb R}\left(\nu_{21}^pB_1(q,t)+\nu_{23}^pB_3(q,t)-\nu_{21}^{ap}B_1^\dagger(q,t)-
\nu_{23}^{ap}B_3^\dagger(q,t)\right)dq\\
\vspace{1mm}
M_3(t)=i{\int}_{\Bbb R}\left(\nu_{31}^pB_1(q,t)+\nu_{32}^pB_2(q,t)-\nu_{31}^{ap}B_1^\dagger(q,t)-
\nu_{32}^{ap}B_2^\dagger(q,t)\right)dq\\
\vspace{1mm}
R_1(t)=\nu_{21}^pp_2(t)+\nu_{21}^{ap}p_2^\dagger(t)+\nu_{31}^pp_3(t)+\nu_{31}^{ap}p_3^\dagger(t)\\
\vspace{1mm}
R_2(t)=\nu_{12}^pp_1(t)+\nu_{12}^{ap}p_1^\dagger(t)+\nu_{32}^pp_3(t)+\nu_{32}^{ap}p_3^\dagger(t)\\
\vspace{1mm}
R_3(t)=\nu_{13}^pp_1(t)+\nu_{13}^{ap}p_1^\dagger(t)+\nu_{23}^pp_2(t)+\nu_{23}^{ap}p_2^\dagger(t).   \label{33}
\end{array}%
\right.
\en

Notice that, if $\nu_{kl}^p=\nu_{kl}^{ap}=0$ for all $k$ and $l$, then $M_j(t)=R_j(t)=0$, $j=1,2,3$, and system (\ref{32}) gives back the one in (\ref{26}). Hence, (\ref{26}) can be seen as a special case of (\ref{32}). Notice also that the system in (\ref{32}) is linear in its dynamical variables, so that an analytic solution can be  found. In fact, its solution can be deduced in a somehow standard way, see \cite{all1,bagbook} and references therein: we first rewrite the equations for $B_j(q,t)$ and $B(q,t)$ in integral form, assume that $\Omega_j(q)$ and $\Omega(q)$ are linear in $q$, $\Omega_j(q)=\Omega_j\,q$ and $\Omega(q)=\Omega\,q$ for some positive constants $\Omega_j$ and $\Omega$, and replace these integral equations in the first three equations in (\ref{32}) for $\dot p_j(t)$. Long but straightforward computations produce the following solution:
\be
P(t)=e^{U\,t}P(0)+\int_0^t e^{U\,(t-t_1)}\,\eta(t_1)\,dt_1,
\label{34}\en
where we have introduced the vectors
$$
P(t)=\left(
       \begin{array}{c}
         p_1(t) \\
         p_2(t) \\
         p_3(t) \\
         p_1^\dagger(t) \\
         p_2^\dagger(t) \\
         p_3^\dagger(t) \\
       \end{array}
     \right), \quad \eta(t)=\left(
                               \begin{array}{c}
                                 \eta_1(t) \\
                                 \eta_2(t) \\
                                 \eta_3(t) \\
                                 \eta_1^\dagger(t) \\
                                 \eta_2^\dagger(t) \\
                                 \eta_3^\dagger(t) \\
                               \end{array}
                             \right),$$ and the symmetric matrix $$U=\left(
\begin{array}{cccccc}
 x_{1,1} & x_{1,2} & x_{1,3}  & y_{1,1} & y_{1,2} & y_{1,3} \\
 x_{1,2} & x_{2,2} & x_{2,3} & \overline{y_{1,2}} & y_{2,2} & y_{2,3} \\
x_{1,3} & x_{2,3} &  x_{3,3} & \overline{y_{1,3}} & \overline{y_{2,3}} &  y_{3,3} \\
 \overline{y_{1,1}} & \overline{y_{1,2}} & \overline{y_{1,3}}  & \overline{x_{1,1}} & \overline{x_{1,2}} & \overline{x_{1,3}} \\
 {y_{1,2}} & \overline{y_{2,2}} & \overline{y_{2,3}} & \overline{x_{1,2}} & \overline{x_{2,2}} & \overline{x_{2,3}} \\
 {y_{1,3}} & {y_{2,3}} & \overline{y_{3,3}} & \overline{x_{1,3}} & \overline{x_{2,3}} & \overline{x_{3,3}} \\
\end{array}
\right).
$$
This is because the set (\ref{32}) can be rewritten, with these definitions, as $\dot P(t)=UP(t)+\eta(t)$.Also,
$$
\left\{
\begin{array}{ll}
x_{1,1}=-i\omega_1-\frac{\pi}{\Omega}\,\tilde\lambda_1^2-\frac{\pi}{\Omega_1}\,\lambda_1^2-\frac{\pi}{\Omega_2}\left((\nu_{12}^p)^2+(\nu_{12}^{ap})^2\right)
-\frac{\pi}{\Omega_3}\left((\nu_{13}^p)^2+(\nu_{13}^{ap})^2\right)\\
\vspace{1mm}
x_{1,2}=-i\mu_{12}^{ex}-\frac{\pi}{\Omega}\,\tilde\lambda_1\tilde\lambda_2-\frac{\pi}{\Omega_1}\,\lambda_1\nu_{21}^p-\frac{\pi}{\Omega_2}\,\lambda_2
\nu_{12}^p-\frac{\pi}{\Omega_3}\left(\nu_{13}^p\nu_{23}^p+\nu_{13}^{ap}\nu_{23}^{ap}\right)\\
\vspace{1mm}
x_{1,3}=-i\mu_{13}^{ex}-\frac{\pi}{\Omega}\,\tilde\lambda_1\tilde\lambda_3-\frac{\pi}{\Omega_1}\,\lambda_1\nu_{31}^p
-\frac{\pi}{\Omega_2}\left(\nu_{12}^p\nu_{32}^p+\nu_{12}^{ap}\nu_{32}^{ap}\right)-\frac{\pi}{\Omega_3}\,\lambda_3
\nu_{13}^p\\
\vspace{1mm}
x_{2,2}=-i\omega_2-\frac{\pi}{\Omega}\,\tilde\lambda_2^2-\frac{\pi}{\Omega_1}\left((\nu_{21}^p)^2+(\nu_{21}^{ap})^2\right)-\frac{\pi}{\Omega_2}\,\lambda_2^2
-\frac{\pi}{\Omega_3}\left((\nu_{23}^p)^2+(\nu_{23}^{ap})^2\right)\\
\vspace{1mm}
x_{2,3}=-i\mu_{23}^{ex}-\frac{\pi}{\Omega}\,\tilde\lambda_2\tilde\lambda_3-\frac{\pi}{\Omega_1}\left(\nu_{21}^p\nu_{31}^p+\nu_{21}^{ap}\nu_{31}^{ap}\right)
-\frac{\pi}{\Omega_2}\,\lambda_2\nu_{32}^p-\frac{\pi}{\Omega_3}\,\lambda_3
\nu_{23}^p\\
\vspace{1mm}
x_{3,3}=-i\omega_3-\frac{\pi}{\Omega}\,\tilde\lambda_3^2-\frac{\pi}{\Omega_1}\left((\nu_{31}^p)^2+(\nu_{31}^{ap})^2\right)
-\frac{\pi}{\Omega_2}\left((\nu_{32}^p)^2+(\nu_{32}^{ap})^2\right)-\frac{\pi}{\Omega_3}\,\lambda_3^2,
\end{array}%
\right.
$$
\vspace{2mm}
$$
\left\{
\begin{array}{ll}
y_{1,1}=-\frac{2\pi}{\Omega_2}\,\nu_{12}^p\nu_{12}^{ap}-\frac{2\pi}{\Omega_3}\,\nu_{13}^p\nu_{13}^{ap}\\
\vspace{1mm}
y_{1,2}=-i\mu_{12}^{coop}-\frac{\pi}{\Omega_1}\,\lambda_1\nu_{21}^{ap}-\frac{\pi}{\Omega_2}\,\lambda_2\nu_{12}^{ap}-\frac{\pi}{\Omega_3}\left(
\nu_{13}^{p}\nu_{23}^{ap}+\nu_{13}^{ap}\nu_{23}^{p}\right)\\
\vspace{1mm}
y_{1,3}=-i\mu_{13}^{coop}-\frac{\pi}{\Omega_1}\,\lambda_1\nu_{31}^{ap}-\frac{\pi}{\Omega_2}\left(
\nu_{12}^{p}\nu_{32}^{ap}+\nu_{12}^{ap}\nu_{32}^{p}\right)-\frac{\pi}{\Omega_3}\lambda_3\nu_{13}^{ap}\\
\vspace{1mm}
y_{2,2}=-\frac{2\pi}{\Omega_1}\,\nu_{21}^p\nu_{21}^{ap}-\frac{2\pi}{\Omega_3}\,\nu_{23}^p\nu_{23}^{ap}\\
\vspace{1mm}
y_{2,3}=-i\mu_{23}^{coop}-\frac{\pi}{\Omega_1}\left(
\nu_{21}^{p}\nu_{31}^{ap}+\nu_{21}^{ap}\nu_{31}^{p}\right)-\frac{\pi}{\Omega_2}\lambda_2\nu_{32}^{ap}-\frac{\pi}{\Omega_3}\lambda_3\nu_{23}^{ap}\\
\vspace{1mm}
y_{3,3}=-\frac{2\pi}{\Omega_1}\,\nu_{31}^p\nu_{31}^{ap}-\frac{2\pi}{\Omega_2}\,\nu_{32}^p\nu_{32}^{ap},
\end{array}%
\right.
$$
and
$$
\left\{
\begin{array}{ll}
\eta_1(t)=i\tilde\lambda_1\beta(t)+i\lambda_1\beta_1(t)+i\nu_{12}^p\beta_2(t)+i\nu_{13}^p\beta_3(t)-i\nu_{12}^{ap}\beta_2^\dagger(t)-i\nu_{13}^{ap}\beta_3^\dagger(t)\\
\vspace{1mm}
\eta_2(t)=i\tilde\lambda_2\beta(t)+i\lambda_2\beta_2(t)+i\nu_{21}^p\beta_1(t)+i\nu_{23}^p\beta_3(t)-i\nu_{21}^{ap}\beta_1^\dagger(t)-i\nu_{23}^{ap}\beta_3^\dagger(t)\\
\vspace{1mm}
\eta_3(t)=i\tilde\lambda_3\beta(t)+i\lambda_3\beta_3(t)+i\nu_{31}^p\beta_1(t)+i\nu_{32}^p\beta_2(t)-i\nu_{31}^{ap}\beta_1^\dagger(t)-i\nu_{32}^{ap}\beta_2^\dagger(t).
\end{array}%
\right.
$$
In these last equations we have further introduced the operators $\beta(t)=\int_{\Bbb R}B(q)e^{-i\Omega q t}dq$ and  $\beta_j(t)=\int_{\Bbb R}B_j(q)e^{-i\Omega_j q t}dq$, $j=1,2,3$.

Let us now call $V_t=e^{Ut}$, which is simply the exponential of a six-by-six matrix, and let us call $(V_t)_{j,l}$ its $(j,l)$-th matrix element. Then the DFs, which are defined as in (\ref{add1}), turn out to be the following functions:
\be
P_j(t)=P_j^X(t)+P_j^Y(t),
\label{35}\en
where
\be
P_j^X(t)=\sum_{l=1}^3\left((V_t)_{3+j,l}(V_t)_{j,3+l}(1-n_l)+(V_t)_{3+j,3+l}(V_t)_{j,l}n_l\right),
\label{36}\en
and
$$
P_j^Y(t)=2\pi\sum_{k,l=1}^3\int_{\Bbb R}dt_1[(V_{t-t_1})_{3+j,k}(V_{t-t_1})_{j,l}q_{k,l}^{(1)}+(V_{t-t_1})_{3+j,k}(V_{t-t_1})_{j,3+l}q_{k,l}^{(2)}+$$
\be
+(V_{t-t_1})_{3+j,3+k}(V_{t-t_1})_{j,l}q_{k,l}^{(3)}+(V_{t-t_1})_{3+j,3+k}(V_{t-t_1})_{j,3+l}q_{k,l}^{(4)}],
\label{37}\en
for $j=1,2,3$. Here, to keep the notation simple, we have still introduced the following quantities:

$$
\left\{
\begin{array}{ll}
q_{1,1}^{(1)}=\frac{\nu_{12}^p\nu_{12}^{ap}}{\Omega_2}+\frac{\nu_{13}^p\nu_{13}^{ap}}{\Omega_3}\\
\vspace{1mm}
q_{1,2}^{(1)}=\frac{1}{\Omega_1}\lambda_1 \nu_{21}^{ap}(1-N_1)+\frac{1}{\Omega_2}\lambda_2 \nu_{12}^{ap}N_2+\frac{1}{\Omega_3}\left(\nu_{13}^{p}\nu_{23}^{ap}(1-N_3)+\nu_{13}^{ap}\nu_{23}^{p}N_3\right)\\
\vspace{1mm}
q_{1,3}^{(1)}=\frac{1}{\Omega_1}\lambda_1 \nu_{31}^{ap}(1-N_1)+\frac{1}{\Omega_2}\left(\nu_{12}^{p}\nu_{32}^{ap}(1-N_2)+\nu_{12}^{ap}\nu_{32}^{p}N_2\right)+
\frac{1}{\Omega_3}\lambda_3 \nu_{13}^{ap}N_3\\
\vspace{1mm}
q_{2,1}^{(1)}=\frac{1}{\Omega_1}\lambda_1 \nu_{21}^{ap}N_1+\frac{1}{\Omega_2}\lambda_2 \nu_{12}^{ap}(1-N_2)+\frac{1}{\Omega_3}\left(\nu_{13}^{p}\nu_{23}^{ap}N_3+\nu_{13}^{ap}\nu_{23}^{p}(1-N_3)\right)\\
\vspace{1mm}
q_{2,2}^{(1)}=\frac{\nu_{21}^p\nu_{21}^{ap}}{\Omega_1}+\frac{\nu_{23}^p\nu_{23}^{ap}}{\Omega_3}\\
\vspace{1mm}
q_{2,3}^{(1)}=\frac{1}{\Omega_1}\left(\nu_{21}^{p}\nu_{31}^{ap}(1-N_1)+\nu_{21}^{ap}\nu_{31}^{p}N_1\right)+\frac{1}{\Omega_2}\lambda_2 \nu_{32}^{ap}(1-N_2)+\frac{1}{\Omega_3}\lambda_3 \nu_{23}^{ap}N_3\\
\vspace{1mm}
q_{3,1}^{(1)}=\frac{1}{\Omega_1}\lambda_1 \nu_{31}^{ap}N_1+\frac{1}{\Omega_2}\left(\nu_{12}^{p}\nu_{32}^{ap}N_2+\nu_{12}^{ap}\nu_{32}^{p}(1-N_2)\right)+
\frac{1}{\Omega_3}\lambda_3 \nu_{13}^{ap}(1-N_3)\\
\vspace{1mm}
q_{3,2}^{(1)}=\frac{1}{\Omega_1}\left(\nu_{21}^{p}\nu_{31}^{ap}N_1+\nu_{21}^{ap}\nu_{31}^{p}(1-N_1)\right)+\frac{1}{\Omega_2}\lambda_2 \nu_{32}^{ap}N_2+\frac{1}{\Omega_3}\lambda_3 \nu_{23}^{ap}(1-N_3)\\
\vspace{1mm}
q_{3,3}^{(1)}=\frac{\nu_{31}^p\nu_{31}^{ap}}{\Omega_1}+\frac{\nu_{32}^p\nu_{32}^{ap}}{\Omega_2},\\
\end{array}%
\right.
$$
and
$$
\left\{
\begin{array}{ll}
q_{1,1}^{(2)}=\frac{1}{\Omega}\tilde\lambda_1^2(1-N)+\frac{1}{\Omega_1}\lambda_1^2(1-N_1)+\frac{1}{\Omega_2}\left((\nu_{12}^p)^2(1-N_2)+(\nu_{12}^{ap})^2N_2\right)+\\\vspace{2mm}
\qquad+\frac{1}{\Omega_3}\left((\nu_{13}^p)^2(1-N_3)+(\nu_{13}^{ap})^2N_3\right)\\
\vspace{1mm}
q_{1,2}^{(2)}=q_{2,1}^{(2)}=\frac{1}{\Omega}\tilde\lambda_1\tilde\lambda_2(1-N)+\frac{1}{\Omega_1}\lambda_1\nu_{21}^p(1-N_1)+\frac{1}{\Omega_2}\lambda_2
\nu_{12}^p(1-N_2)+\\\vspace{2mm}
\qquad+\frac{1}{\Omega_3}\left(\nu_{13}^p\nu_{23}^p(1-N_3)+\nu_{13}^{ap}\nu_{23}^{ap}N_3\right)\\
\vspace{1mm}
q_{1,3}^{(2)}=q_{3,1}^{(2)}=\frac{1}{\Omega}\tilde\lambda_1\tilde\lambda_3(1-N)+\frac{1}{\Omega_1}\lambda_1\nu_{31}^p(1-N_1)+
\frac{1}{\Omega_2}\left(\nu_{12}^p\nu_{32}^p(1-N_2)+\nu_{12}^{ap}\nu_{32}^{ap}N_2\right)+\\
\qquad+\frac{1}{\Omega_3}\lambda_3\nu_{13}^p\nu_{23}^p(1-N_3)\\\vspace{2mm}
\vspace{1mm}
q_{2,2}^{(2)}=\frac{1}{\Omega}\tilde\lambda_2^2(1-N)+\frac{1}{\Omega_1}\left((\nu_{21}^p)^2(1-N_1)+(\nu_{21}^{ap})^2N_1\right)+\frac{1}{\Omega_2}\lambda_2^2(1-N_2)+\\\vspace{2mm}
\qquad+\frac{1}{\Omega_3}\left((\nu_{23}^p)^2(1-N_3)+(\nu_{23}^{ap})^2N_3\right)\\
\vspace{1mm}
q_{2,3}^{(2)}=q_{3,2}^{(2)}=\frac{1}{\Omega}\tilde\lambda_2\tilde\lambda_3(1-N)+\frac{1}{\Omega_1}\left(\nu_{21}^p\nu_{31}^p(1-N_1)+\nu_{21}^{ap}\nu_{31}^{ap}N_1\right)+
\frac{1}{\Omega_2}\lambda_2\nu_{32}^p(1-N_2)+\\
\qquad+\frac{1}{\Omega_3}\lambda_3\nu_{13}^p\nu_{23}^p(1-N_3)\\
\vspace{1mm}
q_{3,3}^{(2)}=\frac{1}{\Omega}\tilde\lambda_3^2(1-N)+\frac{1}{\Omega_1}\left((\nu_{31}^p)^2(1-N_1)+(\nu_{31}^{ap})^2N_1\right)+
\frac{1}{\Omega_2}\left((\nu_{32}^p)^2(1-N_2)+(\nu_{32}^{ap})^2N_2\right)+\\\vspace{2mm}
\qquad+\frac{1}{\Omega_3}\lambda_3^2(1-N_3).
\end{array}%
\right.
$$

As for the $q_{k,l}^{(3)}$, these can be deduced by $q_{k,l}^{(2)}$ simply by replacing $1-N$ with $N$, $N$ with $1-N$, $1-N_j$ with $N_j$ and $N_j$ with $1-N_j$. Hence, for instance, we have
$$
q_{1,1}^{(3)}=\frac{1}{\Omega}\tilde\lambda_1^2N+\frac{1}{\Omega_1}\lambda_1^2N_1+\frac{1}{\Omega_2}\left((\nu_{12}^p)^2N_2+(\nu_{12}^{ap})^2(1-N_2)\right)+$$
$$\qquad+\frac{1}{\Omega_3}\left((\nu_{13}^p)^2N_3+(\nu_{13}^{ap})^2(1-N_3)\right),
$$
and so on.
Finally, simple parity reasons allow us to conclude that $q_{k,l}^{(4)}=q_{l,k}^{(1)}$, for each $k$ and $l$.

We see that $P_j^X(t)$ contains all that refers to the parties, while the coefficients entering in the definition of $P_j^Y(t)$ refer to the electors. Hence formula (\ref{35}) clearly discriminate between the two. Notice also that, as expected, formula (\ref{35}) returns the solution deduced in \cite{all1} when all the $\nu_{kl}^p$ and $\nu_{kl}^{ap}$ are zero. In principle we are now in a position to compute the various DFs for any choice of the parameters and of the initial conditions on the parties and on the electors. A complete analysis of the various scenarios is postponed to a future paper. Here we restrict our analysis to a single particular case, fixing conveniently in the next section the values of the parameters and considering several initial conditions, and plotting the DFs deduced in this way.

\subsection{An explicit example}\label{sectIV}

\begin{figure}[ht]
\begin{center}
\includegraphics[width=0.4\textwidth]{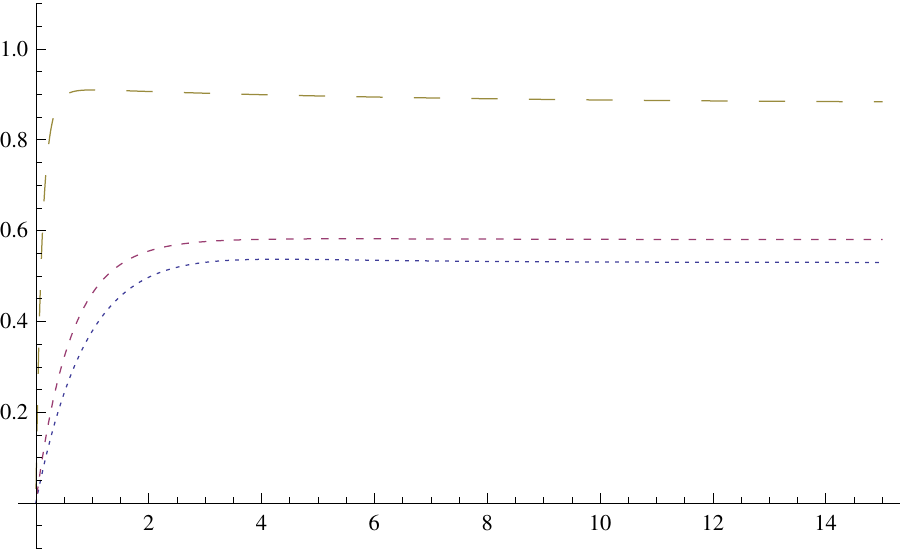}
\includegraphics[width=0.4\textwidth]{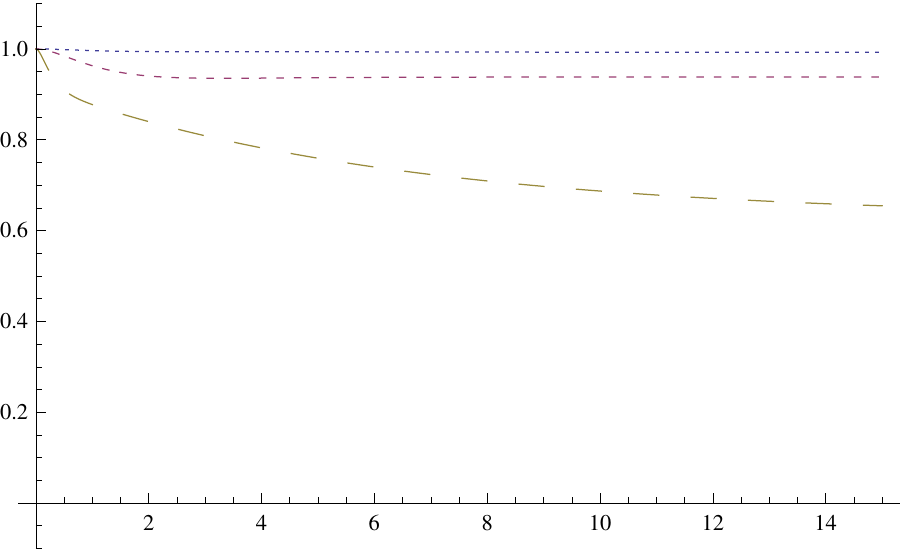}\hfill\\[0pt]
\includegraphics[width=0.4\textwidth]{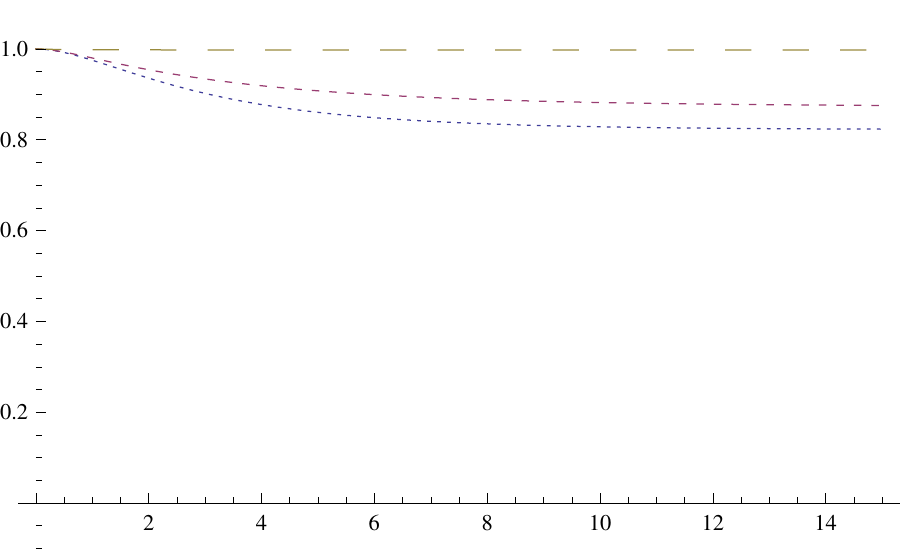}
\end{center}
\caption{{\protect\footnotesize $P_1(t)$ (top left), $P_2(t)$ (top right) and $P_3(t)$ (bottom) for $\mu_{1,2}^{ex}=0.1$, $\mu_{1,3}^{ex}=0.2$, $\mu_{2,3}^{ex}=0.08$ $\mu_{k,l}^{coop}=0$, $\omega_1=\Omega_1=\Omega=0.1$, $\omega_2=\omega_3=\Omega_2=0.2$, $\lambda_1=0.1$, $\lambda_2=0.2$, $\lambda_3=0.05$, $\tilde\lambda_1=0.1$,
$\tilde\lambda_2=\tilde\lambda_3=0$, and $n_1=N_1=0$, $n_2=n_3=N_2=N_3=N=1$.}}
\label{grafico1}
\end{figure}

\begin{figure}[ht]
\begin{center}
\includegraphics[width=0.4\textwidth]{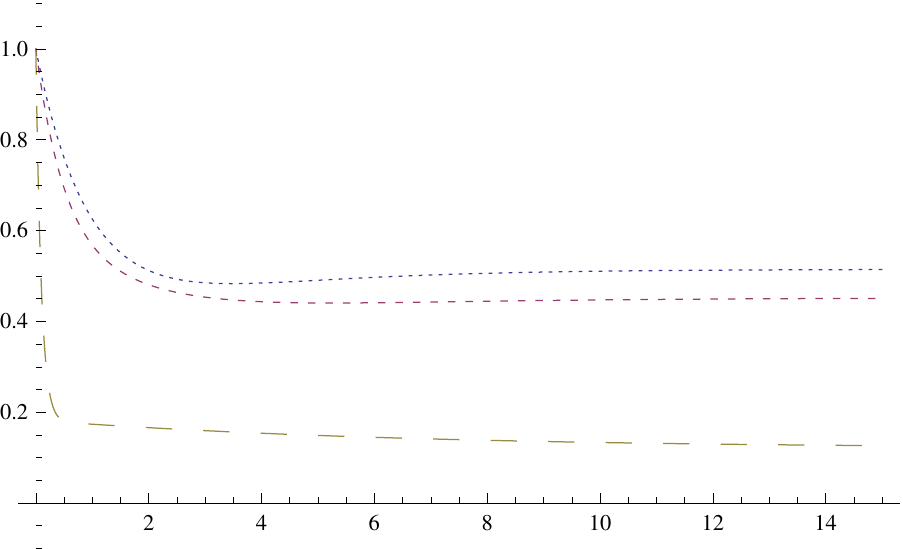}
\includegraphics[width=0.4\textwidth]{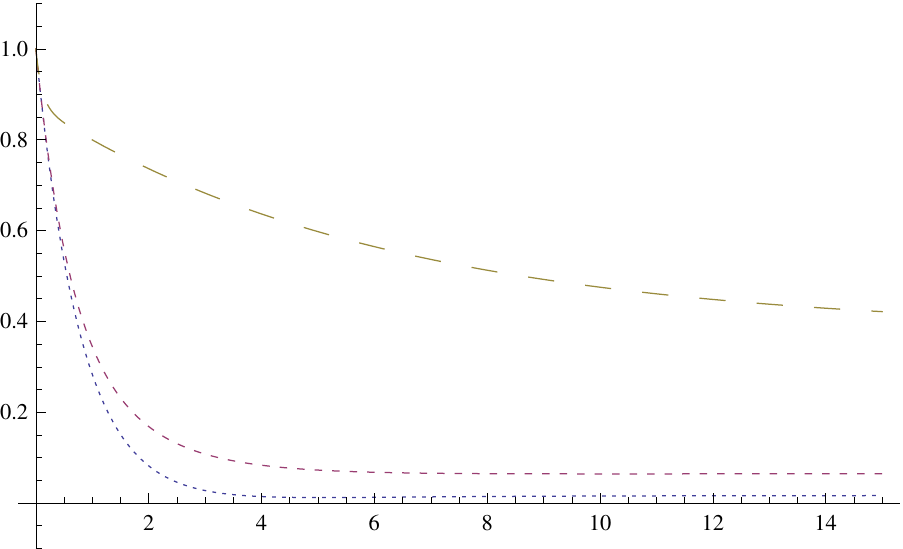}\hfill\\[0pt]
\includegraphics[width=0.4\textwidth]{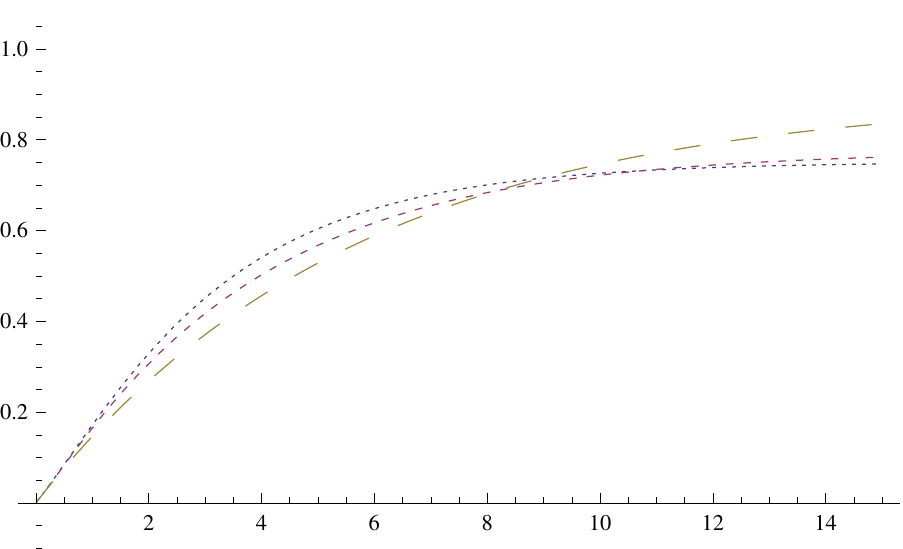}
\end{center}
\caption{{\protect\footnotesize $P_1(t)$ (top left), $P_2(t)$ (top right) and $P_3(t)$ (bottom) for $\mu_{1,2}^{ex}=0.1$, $\mu_{1,3}^{ex}=0.2$, $\mu_{2,3}^{ex}=0.08$ $\mu_{k,l}^{coop}=0$, $\omega_1=\Omega_1=\Omega=0.1$, $\omega_2=\omega_3=\Omega_2=0.2$, $\lambda_1=0.1$, $\lambda_2=0.2$, $\lambda_3=0.05$, $\tilde\lambda_1=0.1$,
$\tilde\lambda_2=\tilde\lambda_3=0$, and $n_3=N_1=N_2=0$, $n_1=n_2=N_3=N=1$.}}
\label{grafico2}
\end{figure}

\begin{figure}[ht]
\begin{center}
\includegraphics[width=0.4\textwidth]{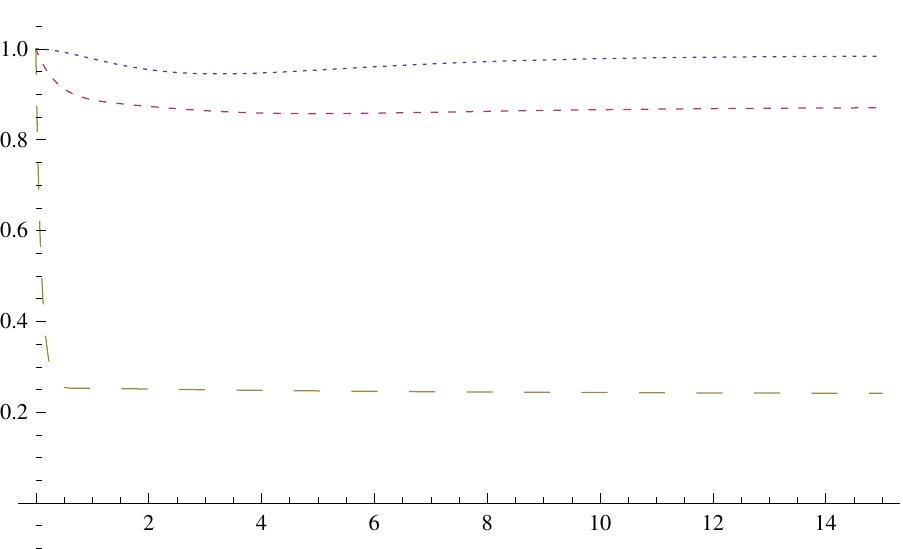}
\includegraphics[width=0.4\textwidth]{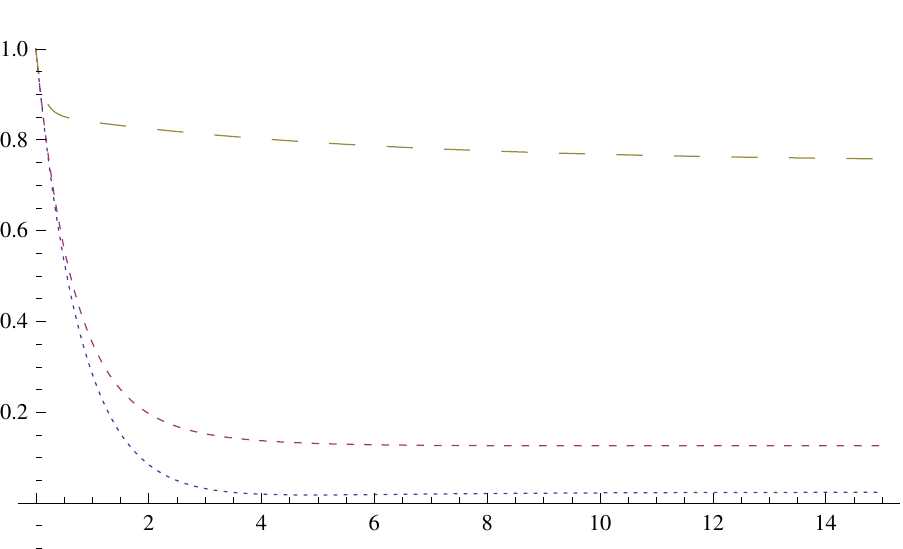}\hfill\\[0pt]
\includegraphics[width=0.4\textwidth]{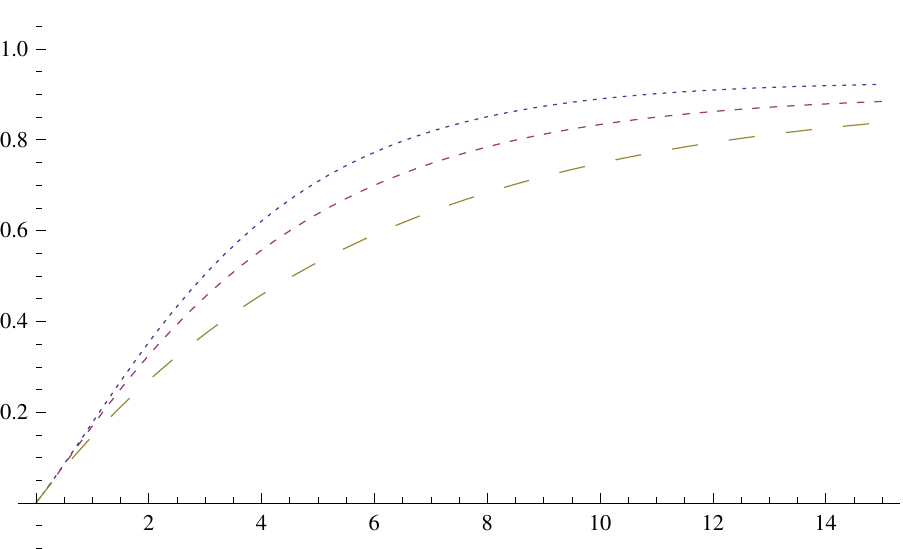}
\end{center}
\caption{{\protect\footnotesize $P_1(t)$ (top left), $P_2(t)$ (top right) and $P_3(t)$ (bottom) for $\mu_{1,2}^{ex}=0.1$, $\mu_{1,3}^{ex}=0.2$, $\mu_{2,3}^{ex}=0.08$ $\mu_{k,l}^{coop}=0$, $\omega_1=\Omega_1=\Omega=0.1$, $\omega_2=\omega_3=\Omega_2=0.2$, $\lambda_1=0.1$, $\lambda_2=0.2$, $\lambda_3=0.05$, $\tilde\lambda_1=0.1$,
$\tilde\lambda_2=\tilde\lambda_3=0$, and $n_3=N_2=0$, $n_1=n_2=N_1=N_3=N=1$.}}
\label{grafico3}
\end{figure}

The situation we will consider here is the first natural extension of the system described in Figure \ref{figscheme}, i.e. a system in which just a single {\em mixed} interaction is allowed. In other words, we will take all the $\nu_{k,l}^{p}$ and $\nu_{k,l}^{ap}$ in $H_{mix}$ equal to zero except $\nu_{12}^p$, which we take different from zero. To simplify further the treatment, we will also take $\tilde\lambda_2=\tilde\lambda_3=0$ and $\mu_{k,l}^{coop}=0$ for all $k$ and $l$. With these choices the only non zero $q_{k,l}^{(j)}$ are the following:

$$
\left\{
\begin{array}{ll}
q_{1,1}^{(2)}=\frac{1}{\Omega}\tilde\lambda_1^2(1-N)+\frac{1}{\Omega_1}\lambda_1^2(1-N_1)+\frac{1}{\Omega_2}(\nu_{12}^p)^2(1-N_2)\\
\vspace{1mm}
q_{1,2}^{(2)}=q_{2,1}^{(2)}=\frac{1}{\Omega_2}\lambda_2
\nu_{12}^p(1-N_2)\\
\vspace{1mm}
q_{2,2}^{(2)}=\frac{1}{\Omega_2}\lambda_2^2(1-N_2)\\
\vspace{1mm}
q_{1,1}^{(3)}=\frac{1}{\Omega}\tilde\lambda_1^2N+\frac{1}{\Omega_1}\lambda_1^2N_1+\frac{1}{\Omega_2}(\nu_{12}^p)^2N_2\\
\vspace{1mm}
q_{2,2}^{(3)}=\frac{1}{\Omega_2}\lambda_2^2N_2\\
\vspace{1mm}
q_{3,3}^{(3)}=\frac{1}{\Omega_3}\lambda_3^2N_3\\
\vspace{1mm}
q_{1,2}^{(3)}=q_{2,1}^{(3)}=\frac{1}{\Omega_2}\lambda_2
\nu_{12}^pN_2
\end{array}%
\right.
$$
The matrix elements of $U$ are also easily found: $y_{k,l}=0$ for all $k$ and $l$, while
$$
\left\{
\begin{array}{ll}
x_{1,1}=-i\omega_1-\frac{\pi}{\Omega}\,\tilde\lambda_1^2-\frac{\pi}{\Omega_1}\,\lambda_1^2-\frac{\pi}{\Omega_2}(\nu_{12}^p)^2\\
\vspace{1mm}
x_{1,2}=-i\mu_{12}^{ex}-\frac{\pi}{\Omega_2}\,\lambda_2
\nu_{12}^p\\
\vspace{1mm}
x_{1,3}=-i\mu_{13}^{ex}\\
\vspace{1mm}
x_{2,2}=-i\omega_2-\frac{\pi}{\Omega_2}\,\lambda_2^2\\
\vspace{1mm}
x_{2,3}=-i\mu_{23}^{ex}\\
\vspace{1mm}
x_{3,3}=-i\omega_3-\frac{\pi}{\Omega_3}\,\lambda_3^2.
\end{array}%
\right.
$$
Then functions $P_j^X(t)$ and $P_j^Y(t)$ in (\ref{36}) and (\ref{37}), and their sums $P_j(t)$, can be easily computed, and the DFs are plotted in Figures \ref{grafico1}-\ref{grafico3} for particular values of the parameters and of the initial conditions.

Here we plot $P_1(t)$, $P_2(t)$ and $P_3(t)$ for $\mu_{1,2}^{ex}=0.1$, $\mu_{1,3}^{ex}=0.2$, $\mu_{2,3}^{ex}=0.08$ $\mu_{k,l}^{coop}=0$, for all $k,l$, $\omega_1=\Omega_1=\Omega=0.1$, $\omega_2=\omega_3=\Omega_2=0.2$, $\lambda_1=0.1$, $\lambda_2=0.2$, $\lambda_3=0.05$, $\tilde\lambda_1=0.1$,
$\tilde\lambda_2=\tilde\lambda_3=0$. The three figures differ for the choice of the initial conditions on the parties and the reservoirs: in Figure \ref{grafico1} we have taken $n_1=N_1=0$, $n_2=n_3=N_2=N_3=N=1$, while in Figure \ref{grafico2} we have $n_3=N_1=N_2=0$, $n_1=n_2=N_3=N=1$ and in Figure \ref{grafico3} $n_3=N_2=0$, $n_1=n_2=N_1=N_3=N=1$. The different lines inside each picture correspond to different values of $\nu_{12}^p$. In particular we have $\nu_{12}^p=0$ for the dotted lines, $\nu_{12}^p=0.1$ for the small dashing lines and $\nu_{12}^p=0.5$ for the large dashing lines. It is evident that $P_3(t)$ is not deeply affected from the presence of this new interaction between $\Pc_1$ and $\Rc_2$. This is not surprising, since $\Pc_3$ is not directly involved in the new term in the Hamiltonian. However, since $\Pc_3$ interacts with $\Pc_1$, and since  $\Pc_1$ can also interact with $\Rc_2$ when $\nu_{12}^p\neq0$, some minor changes are expected and, in fact, this is what we observe in the plots given here for $P_3(t)$, for all choices of initial conditions.

Much more evident is the change in $P_1(t)$ and $P_2(t)$, since $\Pc_1$ and $\Pc_2$ are  directly affected by the new term in the Hamiltonian. In particular we see in these figures that the higher the value of $\nu_{12}^p$, the higher the tendency of $P_1(t)$ to approach the value $N_2$, which describes the initial status of $\Rc_2$: of course, this is due also to the fact that the {\em magnitude} of the interaction between $\Pc_1$ and $\Rc_2$, when increasing $\nu_{12}^p$, becomes more relevant than the interaction between $\Pc_1$ and its own electors, those in $\Rc_1$. This means that $\Pc_1$, for high values of $\nu_{12}^p$, is more interested in the opinion of the electors of $\Pc_2$ than to its own electors!

\section{Conclusions and perspectives}\label{sectV}

In this paper we have considered an extension of a model of political alliances recently proposed by the author. In the framework analyzed here, the dynamical variables are operators and the dynamics is originally defined by means of a self-adjoint Hamiltonian describing the interactions existing among the three political parties and four different sets of electors. With respect to the original model we have considered here the possibility that a given party influences, and is influenced, by the electors of a different party.

While the solution of the dynamical problem is obtained here in its full generality, the explicit examples considered in this paper only refer to a particularly simple choice of the parameters of the Hamiltonian. However, already for this simple choice, the mechanism of the model appears clearly. A deeper analysis of our results, with a detailed study of the role of the parameters of the Hamiltonian, will be the object of a future publication.

\section*{Acknowledgements}

The author acknowledges partial support from Palermo University and from G.N.F.M. of the INdAM. The author also wishes to thank the organizers of the XVIII Wascom Conference for their financial and logistic support during the conference.

\end{document}